\documentclass[preprint,showpacs,preprintnumbers,amsmath,amssymb]{revtex4}
\usepackage{booktabs}
\usepackage{mathrsfs}
\usepackage{epsfig}
\usepackage{graphicx}% Include figure files
\usepackage{dcolumn}% Align table columns on decimal point
\usepackage{bm}% bold math
\usepackage{amsmath}
\usepackage{slashed}       % slashed p
\usepackage{amssymb}
\usepackage{amsfonts}
\usepackage{graphics,float,psfrag}   %%%,axodraw
\usepackage[usenames,dvips]{color}
\usepackage{rotating}

\let\jnfont=\rm
\def\NPB#1,{{\jnfont Nucl.\ Phys.\ B }{\bf #1},}
\def\PLB#1,{{\jnfont Phys.\ Lett.\ B }{\bf #1},}
\def\EPJC#1,{{\jnfont Eur.\ Phys.\ Jour.\ C }{\bf #1},}
\def\PRD#1,{{\jnfont Phys.\ Rev.\ D }{\bf #1},}
\def\PRL#1,{{\jnfont Phys.\ Rev.\ Lett.\ }{\bf #1},}
\def\MPLA#1,{{\jnfont Mod.\ Phys.\ Lett.\ A }{\bf #1},}
\def\JPG#1,{{\jnfont J.\ Phys.\ G}{\bf #1},}
\def\CTP#1,{{\jnfont Commun.\ Theor.\ Phys.\ }{\bf #1},}
\def\ZPC#1,{{\jnfont Z.\ Phys.\ C }{\bf #1},}
\def\JHEP#1,{{\jnfont JHEP \ }{\bf #1},}
\def\lsim{\raise0.3ex\hbox{$<$\kern-0.75em\raise-1.1ex\hbox{$\sim$}}}
\def\gsim{\raise0.3ex\hbox{$>$\kern-0.75em\raise-1.1ex\hbox{$\sim$}}}

\newcommand{\GeV}{~\rm GeV}
\newcommand{\TeV}{~\rm TeV}
\newcommand{\fbm}{{~\rm fb}^{-1}}
\newcommand{\fb}{~\rm fb}

\def\Sign{{\rm Sign}}

\begin{document}
\preprint{\parbox{1.2in}{\noindent arXiv: 1311.6661}}

\title{Higgs Phenomenology in the Minimal Dilaton Model\\
after Run I of the LHC
}

\author{Junjie Cao$^{1,2}$, Yangle He$^1$, Peiwen Wu$^3$, Mengchao Zhang$^3$, Jingya Zhu$^3$}

\affiliation{
  $^1$  Department of Physics,
        Henan Normal University, Xinxiang 453007, China \\
  $^2$ Center for High Energy Physics, Peking University,
       Beijing 100871, China \\
  $^3$ State Key Laboratory of Theoretical Physics,
      Institute of Theoretical Physics, Academia Sinica, Beijing 100190, China
      \vspace{0.36cm}}

\begin{abstract}
The Minimal Dilaton Model (MDM) extends the Standard Model (SM) by a singlet scalar,
which can be viewed as a linear realization of general dilaton field.
This new scalar field mixes with the SM Higgs field to form two mass eigenstates with one of
them corresponding to the $125\GeV$ SM-like Higgs boson reported by the LHC experiments.
In this work, under various theoretical and experimental constrains, we perform fits to the latest Higgs data and then
investigate the phenomenology of Higgs boson in both the heavy dilaton scenario and the light dilaton scenario of the MDM.
We find that: (i) If one considers the ATLAS and CMS data separately, the MDM can explain each of them well,
but refer to different parameter space due to the apparent difference in the two sets of data.
If one considers the combined data of the LHC and Tevatron, however, the explanation given by the MDM is not much better than the SM,
and the dilaton component in the 125-GeV Higgs is less than about $20\%$ at $2\sigma$ level.
(ii) The current Higgs data have stronger constrains on the light dilaton scenario than on the heavy dilaton scenario.
(iii) The heavy dilaton scenario can produce a Higgs triple self coupling much larger than the SM value, and
thus a significantly enhanced Higgs pair cross section at hadron colliders.
With a luminosity of $100\fbm$ ($10\fbm$) at the 14-TeV LHC, a heavy dilaton of $400\GeV$ ($500\GeV$) can be examined.
(iv) In the light dilaton scenario, the Higgs exotic branching ratio can reach $43\%$ (60\%) at $2\sigma$ ($3\sigma$) level
when considering only the CMS data, which may be detected at the 14-TeV LHC with a luminosity of $300\fbm$ and the Higgs Factory.

\end{abstract}


\maketitle

\section{Introduction}
Based on about $25\fbm$ data collected at 7-TeV and 8-TeV LHC, the ATLAS and
CMS collaborations have further corroborated the existence of a new boson with a
local statistical significance reaching $9 \sigma$ and more than $7\sigma$, respectively
\cite{1207-a-com,1207-c-com,1303-atlas-Moriond,1303-c-com}.
So far the mass of the boson is rather precisely determined to be around 125 GeV,
and its other properties, albeit with large experimental uncertainties,
agree with those of the Higgs boson predicted by the Standard Model (SM)
\cite{1303-c-com,1303-a-com}. Nevertheless, the deficiencies of the SM itself motivate
the interpretation of the Higgs-like boson in new physics frameworks such as low energy
supersymmetric models \cite{SUSY-our, Light-1309}, and as shown
by numerous studies, fits to the Higgs data in new physics models can be as good as that in the SM.

Among the new physics interpretations of the Higgs-like boson, dilaton is another attractive one.
This particle arises from a strong interaction theory with approximate scale invariance at a certain high energy scale.
Then the breakdown of the invariance triggers the electroweak symmetry breaking, and the dilaton
as the Nambu-Goldstone particle of the broken invariance can be naturally light in comparison with
the high energy scale. In this framework, the whole SM sector is usually assumed to be a part of the scale
invariance maintained at the UV scale, and all the fermions and gauge bosons of the SM are composite particles
at weak scale. After such treatment, the couplings of the linearized dilaton field $S$ to the SM fields take the
form \cite{dilaton-before}
\begin{eqnarray}
{\cal{L}} = \frac{S}{f} T^\mu_\mu,
\end{eqnarray}
with $f$ denoting the dilaton decay constant and $T^\mu_\mu$ representing the trace of the energy-momentum
tensor of the SM. Through this term, the dilaton couples directly to the violation of the scale invariance 
in the SM, i.e., to the fermions and $W$, $Z$ bosons with strength proportional to their masses, and thus mimics 
the behavior of the SM Higgs boson at Run I of the LHC. Since the first hint of the Higgs-like boson at the LHC 
was uncovered at the end of 2011, the compatibility of the dilaton with the data has been extensively
discussed \cite{dilaton, hfit-first, hfit-dilaton, hfit-1303, dilaton-tc}.
For example, in \cite{hfit-first, hfit-dilaton, hfit-1303} the traditional dilaton models were compared with
models including Higgs boson, and the techni-dilaton was shown to be able to explain the signals well \cite{dilaton-tc}.
The Higgs-dilaton was also used to solve cosmological problems such as inflation and dark energy \cite{dilaton-th}.

In this work, we concentrate on the Minimal Dilaton Model (MDM), which is actually a minimal effective Lagrangian at weak
scale describing the breaking of a UV strong dynamics with scale invariance \cite{min-dilaton1, min-dilaton2}.
This model is motivated by topcolor theory \cite{Hill}, and it introduces one massive vector-like fermion with
the same quantum number as right-handed top quark. The mass of this top partner represents the scale of the dynamical sector,
to which the dilaton naturally couples in order to recover the scale invariance. With such setting, top quark as the mass eigenstate
couples to the strong dynamics by its mixing with the partner. Moreover, unlike the traditional dilaton model,
the SM except for the Higgs field acts as a spectator of the dynamics, and consequently the dilaton does not couple directly
the fermions and the $W$, $Z$ bosons in the SM. In this sense, the dilaton is equivalent to an electroweak gauge singlet field.

In the Minimal Dilaton Model, the SM Higgs field and the dilaton field mix to form two CP-even mass eigenstates.
Hereafter we call the eigenstate with the Higgs field as its dominant component Higgs particle, and the other one dilaton.
The property of the Higgs boson may deviate significantly from that of the SM Higgs boson due to the mixing effect and also due to
its interactions with the dilaton. Noting that the di-photon rate of the Higgs-like boson reported by the CMS collaboration
at the Rencontres de Moriond 2013 differs greatly from its previous publications, we in the following first update the fit
in \cite{min-dilaton1, min-dilaton2} by using the latest Higgs data. Then we consider the phenomenology of the Higgs boson
at the LHC. We are particularly interested in the following two scenarios:
\begin{itemize}
\item Heavy dilaton scenario, where the dilaton is heavier than the Higgs boson. In this scenario, the triple Higgs coupling
may be potentially large and consequently, the Higgs pair production rate at the LHC can be greatly enhanced.
\item Light dilaton scenario, where the dilaton is lighter than half the Higgs boson mass. In this case, the Higgs boson
may decay into the dilaton pair with a sizeable branching ratio.
\end{itemize}

The paper is organized as follows. In Section II, we briefly review the Minimal Dilaton Model.
In Section III, we concentrate on the heavy dilaton scenario and scan through the MDM parameter space
by considering various theoretical and experimental constraints. For the surviving samples, we perform fits to the
Higgs data and study the Higgs pair production and its detection at the LHC. In Section IV, we turn to investigate the
light dilaton scenario in a similar way, but pay particular attention to the exotic
decay of the Higgs boson into dilaton pair. Finally, we draw our conclusions in Section V.

\section{the Minimal Dilaton Model}
The Minimal Dilaton Model extends the SM by one gauge singlet scalar field $S$ which represents a linearized dilaton
field, and one fermion field $T$ with the same quantum number as the right-handed top quark which is usually called the top quark partner.
Its effective Lagrangian can be written as \cite{min-dilaton1,min-dilaton2}
\begin{eqnarray}
\mathcal L &=
	&	\mathcal L_\text{SM}
		-{1\over2}\partial_\mu S\partial^\mu S
		-\tilde V(S,H)\nonumber\\
	&&   -\overline T\left(\slashed{D}+\frac{M}{f}S\right)T
		-\left[y'\overline T_R(q_{3L}\cdot H)+\text{h.c.}\right],  \label{MDM}
\end{eqnarray}
where $q_{3L}$ is the $SU(2)_L$ left-handed quark doublet of the third generation,
$M$ the scale of the strong dynamics, and $\mathcal L_\text{SM}$ is the SM
Lagrangian without Higgs potential. In Eq.(\ref{MDM}), $\tilde V(S,H)$ as scalar potential describes the interactions of $S$ with the SM Higgs field $H$. Its general expression is given by
\begin{eqnarray}
\tilde V(S,H)
	&=	{m_S^2\over2}S^2+{\lambda_S\over 4!}S^4+{\kappa\over2}S^2\left|H\right|^2
		+m_H^2\left|H\right|^2+{\lambda_H\over4}\left|H\right|^4,
		\label{potential}
\end{eqnarray}
where $m_S$, $\lambda_S$, $\kappa$, $m_H$ and $\lambda_H$ are all free real parameters. With such potential,
the field $S$ and $H$ will mix to form two CP-even mass eigenstates, i.e. the Higgs boson $h$ and the dilaton $s$,
and the mixing angle $\theta_S$ is defined by
\begin{eqnarray}
  H^0 &=& \frac{1}{\sqrt{2}} (v+h\cos\theta_S-s\sin\theta_S), \nonumber\\
  S &=& f+h\sin\theta_S+s\cos\theta_S,
\end{eqnarray}
with $f$ and $v/\sqrt{2}=174$ GeV denoting the vacuum expectation values (vevs) of $S$ and $H$, respectively.
Detailed studies indicate that if $h$ instead of $s$ corresponds to the newly discovered boson, a much lower $\chi^2$ can be obtained
in the fits to the Higgs data. So in our discussion we fix $m_h= 125.6\GeV$ which is the combined mass
value of the two collaborations \cite{hfit-1303}. As for the potential, it is more convenient to use
$f$, $v$, $\theta_S$, $m_h$ and $m_s$ as input parameters. In this case, $\kappa$, $\lambda_H$ and
$\lambda_S$ can be re-expressed as
\begin{eqnarray}
\kappa &=& \frac{|m_h^2 - m_s^2|}{2 f v} |\sin2\theta_S| = \frac{|m_h^2-m_s^2|}{v^2} \frac{|\eta \tan \theta_S|}{1+\tan^2 \theta_S} \geq0, \nonumber\\
\lambda_H &=& \frac{|m_h^2 - m_s^2|}{v^2} ~\Big[~\big|\frac{m_h^2 + m_s^2}{m_h^2 - m_s^2} \big| +\Sign{(\sin2\theta_S)} \cos2\theta_S \Big],
\nonumber\\
\lambda_S &=& \frac{3|m_h^2 - m_s^2|}{2f^2} ~\Big[~\big|\frac{m_h^2 + m_s^2}{m_h^2 - m_s^2} \big | -\Sign{(\sin2\theta_S)} \cos2\theta_S \Big],
\label{lambda}
\end{eqnarray}
where $\Sign{(\sin2\theta_S)}$ denotes the sign of $\sin2\theta_S$.
Then the triple Higgs self coupling normalized to its SM value
and the Higgs coupling to a dilaton pair
are given by
\begin{eqnarray}
C_{hhh}/SM = \frac{v^2}{3m_h^2}
&\big [&~\frac{3}{2} \lambda_H\cos^3\theta_S +\lambda_S\eta^{-1}\sin^3\theta_S \nonumber\\
&&+3\kappa(\cos\theta_S\sin^2\theta_S +\eta^{-1}\cos^2\theta_S\sin\theta_S) ~\big],
\label{selfcoup}
\end{eqnarray}
\begin{eqnarray}
C_{hss} = v &\big[&
   \kappa (\cos^3\theta_S
 + \eta^{-1} \sin^3\theta_S)
 + (\frac{3}{2} \lambda_H-2\kappa) \cos\theta_S \sin^2\theta_S
 \nonumber\\
 && + \eta^{-1} (\lambda_S-2\kappa) \cos^2\theta_S \sin\theta_S ~\big].
\label{hsscoup}
\end{eqnarray}
In the above expressions, we define \cite{min-dilaton1,min-dilaton2}
\begin{eqnarray}
  \eta &\equiv& \frac{v}{f} N_T ,
\end{eqnarray}
where $N_T$ denotes the number of $T$ field and is set to 1 for the MDM. Note that $C_{hhh}/SM$ may be much larger than 1 in the case of heavy dilaton scenario.

Similar to Eq.(\ref{lambda}), one may also define the top quark mixing angle in terms of mass eigenstates $t$ and $t^\prime$ as
\begin{eqnarray}
q_{3L}^u &=& \cos \theta_L t_L + \sin \theta_L t^\prime_L, \nonumber \\
T_L &=& - \sin \theta_L t_L + \cos \theta_L t^\prime_L.
\end{eqnarray}
Then under the conditions $m_{t'}\gg m_t$ and $\tan\theta_L\ll m_{t'}/m_t$, the
normalized couplings of $h$ and $s$ are given by \cite{min-dilaton1,min-dilaton2}
\begin{eqnarray}
C_{hVV}/SM	&=&	C_{hff}/SM	= \cos\theta_S,\nonumber\\
C_{htt}/SM	&=& \cos^2\theta_L\cos\theta_S +\eta\sin^2\theta_L\sin\theta_S, \nonumber\\
C_{ht't'}/SM &=& \sin^2\theta_L\cos\theta_S +\eta\cos^2\theta_L\sin\theta_S, \nonumber\\
C_{hgg}/SM	&=&	\cos\theta_S +\eta\sin\theta_S,\nonumber\\
C_{h\gamma\gamma}/SM &=&\cos\theta_S -0.27\times\eta\sin\theta_S,
\label{scoupling}
\end{eqnarray}
and
\begin{eqnarray}
C_{sVV}/SM	&=& C_{sff}/SM	=-\sin\theta_S,\nonumber\\
C_{stt}/SM	&=& -\cos^2\theta_L\sin\theta_S +\eta\sin^2\theta_L\cos\theta_S,
\nonumber\\
C_{st't'}/SM &=& -\sin^2\theta_L\sin\theta_S +\eta\cos^2\theta_L\cos\theta_S,
\nonumber\\
C_{sgg}/SM	&=&	[A_b \cos\theta_S
+A_t \times (-\cos^2\theta_L\sin\theta_S +\eta\sin^2\theta_L\cos\theta_S)
\nonumber\\
&&+A_{t'} \times (-\sin^2\theta_L\sin\theta_S +\eta\cos^2\theta_L\cos\theta_S)] /(A_t+A_b),
 \nonumber\\
C_{s\gamma\gamma}/SM &=&
[(A_W+\frac{1}{3}A_b) \times \cos\theta_S
+\frac{4}{3}A_t \times (-\cos^2\theta_L\sin\theta_S +\eta\sin^2\theta_L\cos\theta_S)
\nonumber\\
&&+\frac{4}{3}A_{t'} \times (-\sin^2\theta_L\sin\theta_S +\eta\cos^2\theta_L\cos\theta_S) ]/[A_W+\frac{4}{3}A_t+\frac{1}{3}A_b)].
\label{hcoupling}
\end{eqnarray}
where $V$ denotes either $W^{\pm}$ or $Z$ boson, $f$ the fermions except for top quark,
and $A_i$ is the loop function presented in \cite{hsm-rv} with particle $i$ running in the loop.

\section{Higgs phenomenology in heavy dilaton scenario}

In this section, we first scan over the parameter space of the Minimal Dilaton Model in the heavy dilaton scenario
under various constraints. Then for the surviving samples we investigate the features of $h$, such as
its couplings to SM particles, $s$ and itself. Before our scan, we clarify the following facts
\begin{itemize}
\item Firstly, since the property of the dilaton in the MDM differs greatly from that of the SM Higgs boson, its mass may vary
from several GeV to several hundred GeV without conflicting with LEP and LHC data in searching for Higgs boson.
In fact, these data actually require the mass of the SM Higgs boson to be above $114 \GeV$ and outside the region
$127-710 \GeV$, respectively \cite{1304-c-exclude}.
\item Secondly, since we are more interested in new physics at low energy, we take $0 < \eta^{-1} \leq 10$
with $\eta^{-1}\equiv f/v$ in our study and pay special attention to the case $\eta^{-1} \geq 1$.
\item Thirdly, although in principle $\theta_S$ may vary from $-\pi/2$ to $\pi/2$, the Higgs data have required it to be around zero so that
$h$ is mainly responsible for the electroweak symmetry breaking. In practice, requiring $|\tan\theta_S| \leq 2$ will suffice.
\item Finally, we note that $t^\prime$ mass has been limited. For example, the ATLAS and CMS collaborations have set
a lower bound of $656 \GeV$ and $685 \GeV$ in the search for the
top partner at the LHC \cite{ex-top-partner} at $95\%$ confidence level, respectively. The requirement of perturbativity, however, sets an upper bound of about $3400 \GeV$ for the $t^\prime$ mass \cite{min-dilaton1}.
\end{itemize}
In summary, for the heavy dilaton scenario we scan the following parameter space:
\begin{eqnarray}
  && 0< \eta^{-1} <10, ~~ |\tan\theta_S|<2, ~~ 0< \sin\theta_L <1, \nonumber\\
  && 130~{\rm GeV} <m_s< 1000~{\rm GeV}, ~~ 700~{\rm GeV} <m_{t'}< 3000~{\rm GeV}.
\end{eqnarray}
In our scan, we consider following constraints:
\begin{itemize}
\item[(1)] The vacuum stability of the scalar potential, which requires $\lambda_H \lambda_S -6\kappa^2 >0$ \cite{min-dilaton1}.
  We analytically checked that this inequality is sufficient to guarantee $\langle H \rangle = v/\sqrt{2}$, $\langle S \rangle = f$ corresponding to
  the global minimum of the potential.
\item[(2)] Absence of the Landau pole for dimensionless couplings $\lambda_H$, $\lambda_S$ and $\kappa$ below $1 \TeV$.
 For each sample in our scan, we solve numerically following renormalization group equations (RGE)  \cite{SMSRGE}
\begin{eqnarray}
 && 16 \pi^2 \mu \frac{d}{d \mu} \lambda_H = 6 \lambda_H^2 +2\kappa^2,  \nonumber\\
 && 16 \pi^2 \mu \frac{d}{d \mu} \lambda_S = 3 \lambda_S^2 +12\kappa^2,  \nonumber\\
 && 16 \pi^2 \mu \frac{d}{d \mu} \kappa = \kappa (3\lambda_H + \lambda_S) +4\kappa^2,
\end{eqnarray}
and require each of the three couplings less than 1000 before reaching $\mu=1\TeV$. In building the RGE,
we have properly considered the normalization factors of the fields, and neglect the effects of the
gauge and Yukawa couplings since we are only interested in the case of large $\lambda_H$, $\lambda_S$
and $\kappa$.

The Landau pole constraint can set upper bounds on the couplings, and the higher scale we choose to impose it,
the tighter the constraint becomes.

\item[(3)] Experimental constraints from the LEP, Tevatron and LHC search for Higgs-like particle.
We implement these constraints with the package $\textsf{HiggsBounds-4.0.0}$ \cite{HiggsBounds}.
\item[(4)] Experimental constraints from the electroweak precision data. We calculate the Peskin-Takeuchi $S$ and $T$ parameters \cite{STU} with the formulae presented in \cite{min-dilaton1}, and construct $\chi^2_{ST}$ with the following experimental fit results \cite{EWPD-fit}:
\begin{eqnarray}
  S=0.03\pm0.10, ~~ T=0.05\pm0.12, ~~ \rho_{ST}=0.89.
\end{eqnarray}
We keep samples with $\chi^2_{ST}<6.18$ for further study. We do not consider the constraints from $V_{tb}$ and $R_b$ since they are weaker than the $S,T$ parameters \cite{min-dilaton1}.

\item[(5)] Experimental constraints from the Higgs data after the Rencontres de Moriond 2013. These data include the exclusive signal rates for
$\gamma \gamma$, $ZZ^\ast$, $W W^\ast$, $b\bar{b}$ and $\tau \bar{\tau}$ channels,
and their explicit values are summarized in Fig.2 of ref.\cite{1303-a-com} for the ATLAS results,
in Fig.4 of ref.\cite{1303-c-com} for the CMS results and in Fig.15 of ref.\cite{1303-t-com}
for the CDF+D0 results. Similar to our previous works \cite{hfit-our},
we perform fits to the data using the method first introduced in \cite{hfit-first} and consider correlations of the data as done in \cite{1212-Gunion, 1307-Gunion}. As for branching ratios of various decay channels for different Higgs boson masses in the SM, we used the results in \cite{h-incl-handbook}.
\end{itemize}

Moreover, since the latest di-photon rate reported by the CMS collaboration ($0.77 \pm 0.27$ \cite{1303-c-com})
is much smaller than that of the ATLAS group ($1.6\pm 0.3$ \cite{1303-a-com}),
we perform three independent fits by using only the ATLAS data (9 sets), only the CMS data (9 sets) and the
combined data (22 sets) including ATLAS, CMS, and CDF+D0.
We checked that $\chi^2$ in the SM are $10.64$, $4.78$ and $18.79$ for the three fits, and
$\chi^2_{min}$ in the MDM are $8.32$, $2.57$ and $18.66$, respectively. The fact that
$\chi^2_{min} < \chi^2_{SM}$ reflects that the MDM is more adaptable to the data than the SM.

In our discussion, we are particularly interested in two types of samples, i.e.,
$\Delta\chi^2\le 2.3$ and $2.3 \leq \Delta\chi^2\leq 6.18$ with $\Delta\chi^2=\chi^2 - \chi^2_{min}$. These two sets of samples correspond to the $68\%$
and $95\%$ confidence level regions in any two dimensional plane of the model parameters when explaining
the Higgs data \cite{1212-Gunion, 1307-Gunion}. Hereafter we call them $1\sigma$ and $2\sigma$
samples, respectively.

\subsection{The heavy dilaton scenario confronted with the current Higgs data}
%%Fig.1 %%%%%%%%%%%%%%%%%%%%
\begin{figure}[]
\includegraphics[width=15.0cm]{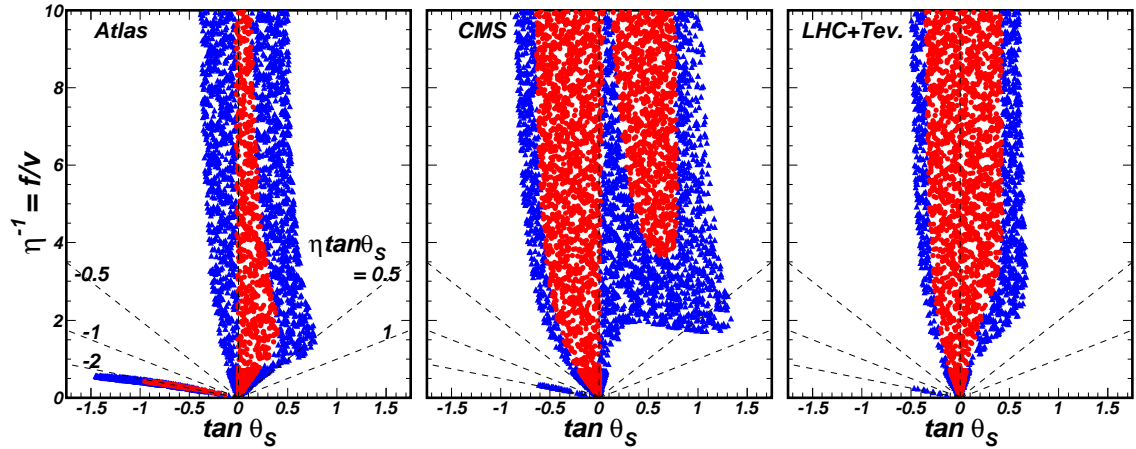}\vspace{0.1cm}
\vspace*{-0.5cm}
\caption{The scatter plots of the samples passing the constraints of the ATLAS data (left), the CMS data (middle) and the LHC+Tevatron data
(right), respectively, projected on the plane of $\eta^{-1}\equiv f/v$ versus $\tan\theta_S$.
All the samples have survived the constraints listed in the text with the red bullets and blue triangles corresponding to
samples further satisfying $\Delta\chi^2 <2.30$  and $2.30< \Delta\chi^2 <6.18$ respectively. Six dashed lines of $\tan\theta_S=0$
and $\eta\tan\theta_S=\pm0.5,\ \pm1,\ -2$ are also shown for convenience of later analysis.
}
\label{fig1}
\end{figure}
%%%%%%%%%%%%%%%%%%%%

In Fig.\ref{fig1}, we project the $1\sigma$ (red bullets) and $2\sigma$ (blue triangles) samples passing various experimental constraints on the plane of $\eta^{-1}\equiv f/v$ versus $\tan\theta_S$.
From this figure we can see that:
\begin{itemize}
\item The latest Higgs data is very powerful in constraining the MDM parameter space. For the three different fits, most of the
samples are confined in the narrow region of $|\tan\theta_S|<1$ and $|\eta\tan\theta_S|<0.5$ or the region of
$\eta^{-1} \lesssim 1$ and $\eta\tan\theta_S\simeq-2$. 
Furthermore, detailed analysis of our results indicates that, 
due to the Landau pole constraint, $\lambda_H$, $\lambda_S$ and $\kappa$ are upper bounded by about 18, 35 and 4.5 respectively.
The relatively smallness of $\kappa$ can also be inferred from Eq.(\ref{lambda}).
  \item
The region of the $1\sigma$ samples for the ATLAS data is approximately complementary to that for the CMS data, which reflects the conflict between the results of the two collaborations.
  \item
We checked that, the minimal $\chi^2$ for the ATLAS data comes from the sample around $\eta\tan\theta_S\simeq-2$ and $\tan\theta_S\simeq-0.2$. According to $(C_{h\gamma\gamma}/SM)/(C_{hff}/SM)=1-0.27\eta\tan\theta_S$ suggested by Eq.(\ref{scoupling}), we can see that the $\chi^2_{\rm min}$ sample for the ATLAS data has an enhanced di-photon signal rate of around 1.55, which is the most attractive feature of the ATLAS data.
  \item
For the combined data, most of the samples are located within $|\tan\theta_S|\lesssim0.5$ or $|\sin\theta_S|\lesssim0.45$, which indicates
that the dilaton/singlet component in the Higgs boson should be less than about $20\%$.
\end{itemize}

%%Fig.2 %%%%%%%%%%%%%%%%%%%%
\begin{figure}[]
\includegraphics[width=15.0cm]{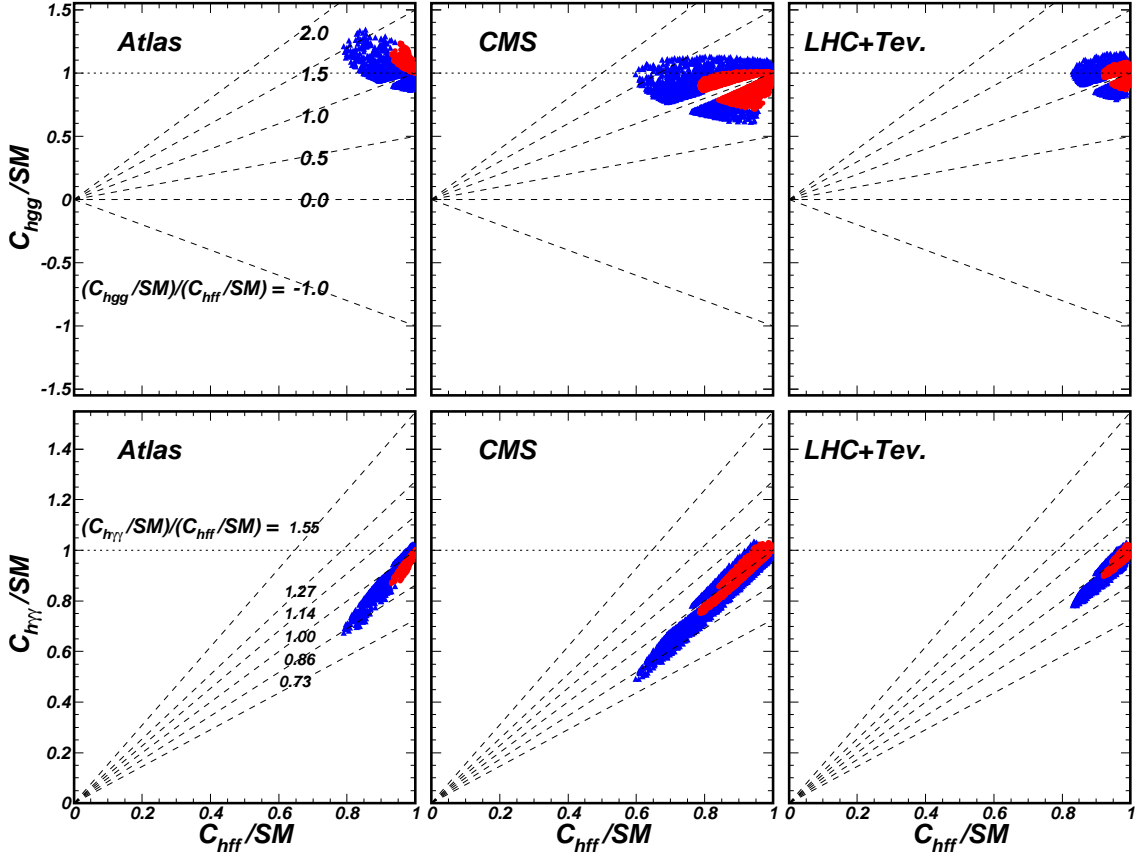}\vspace{0.1cm}
\vspace*{-0.5cm}
\caption{Same as Fig.\ref{fig1}, but only for samples with $\eta^{-1}\equiv f/v > 1 $ and projected on the plane of $C_{hgg}/SM$ (upper) and  $C_{h\gamma\gamma}/SM$ (lower) versus $C_{hff}/SM=C_{hVV}/SM$, respectively. $C_{hXX}/SM$ denotes the normalized 125-GeV Higgs coupling to SM gluon, photon, light fermion and massive vector boson.
The SM point $(C_{hXX}/SM)=1$ corresponds to $\theta_S=0$. According to Eq.(\ref{scoupling}), dashed lines of $(C_{hgg}/SM)/(C_{hff}/SM)=1,\ 1.5,\ 0.5,\ 2,\ 0, -1$ or $(C_{h\gamma\gamma}/SM)/(C_{hff}/SM)=1,\ 0.86,\ 1.14,\ 0.73,\ 1.27,\ 1.55$ correspond to $\eta\tan\theta_S=0,\ \pm0.5,\ \pm1, -2$ in Fig.\ref{fig1}, respectively.
}
\label{fig2}
\end{figure}
%%%%%%%%%%%%%%%%%%%%
In Fig.\ref{fig2}, we only consider the samples with $\eta^{-1} > 1$ and show the normalized Higgs couplings to SM gluon,
photon, light fermion and massive vector boson ($C_{hVV}/SM=C_{hff}/SM$). Since in the MDM,  $C_{hff}$ and $C_{hVV}$ contribute to about $90\%$ of
the total decay width for the SM-like Higgs boson and $C_{hgg}$ dominates the Higgs production rate at the LHC,
the normalized $XX$ signal rate is roughly proportional to $(C_{hgg}/SM)^2/(C_{hff}/SM)^2 \times (C_{hXX}/SM)^2$.
Then combining Fig.\ref{fig1}, Fig.\ref{fig2} and Eq.(\ref{scoupling}), one can learn that:
\begin{itemize}
\item The six dashed lines of $(C_{hgg}/SM)/(C_{hff}/SM)=$1, 1.5, 0.5, 2, 0, -1 in the upper plane and
$(C_{h\gamma\gamma}/SM)/(C_{hff}/SM)=$1,\ 0.86,\ 1.14,\ 0.73,\ 1.27,\ 1.55 in the lower plane actually
correspond to dashed lines of $\eta\tan\theta_S=0,\ \pm0.5,\ \pm1, \ -2$ in Fig.\ref{fig1}, respectively.
Thus the missing parts around $(C_{hgg}/SM)/(C_{hff}/SM)=1$ is actually a consequence of the upper limit of $\eta^{-1}\leq 10$.
\item Most of the samples for the ATLAS data are characterized by $(C_{hgg}/SM)/(C_{hff}/SM)$ $\gtrsim1$.
This is because some of the ATLAS inclusive signals are enhanced compared with their SM values, especially for the di-photon channel.
\item Most of the $1\sigma$ samples for the the CMS data satisfy $(C_{hgg}/SM)\lesssim 1$ and
$(C_{h\gamma\gamma}/SM)/(C_{hff}/SM)\simeq 1$, because most of the CMS inclusive signal rates, including the
latest di-photon rate, are smaller than $1$.
\item When considering the ATLAS and CMS data separately, the MDM can explain each of them well but referring
to different parameter space due to the apparent difference in the two sets of data.
When considering the combined data, however, the MDM explanation is not much better than the SM, and
the 125-GeV Higgs couplings to light fermions and vector bosons are very close to the SM values, especially for the $1\sigma$ samples.
\end{itemize}

%%Fig.3 %%%%%%%%%%%%%%%%%%%%
\begin{figure}[]
\includegraphics[width=15.0cm]{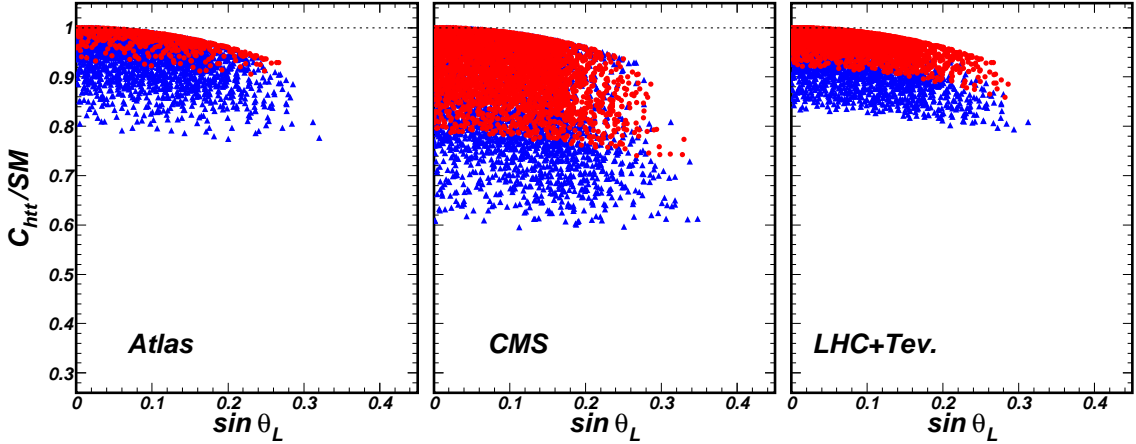}\vspace{0.1cm}
\vspace*{-0.5cm}
\caption{Same as Fig.\ref{fig2}, but projected on the plane of $C_{htt}/SM$ versus $\sin\theta_L$.
}
\label{fig3}
\end{figure}
%%%%%%%%%%%%%%%%%%%%
In above analysis, we checked that the Higgs data have very little constraint on $\theta_L$ and $m_{t'}$.
We also checked that although the EWPD has no constraint on $\eta^{-1}$ and $\tan\theta_S$, it is very powerful
in constraining $\sin\theta_L$ and the coupling $C_{ht\bar{t}}$. The latter is shown in Fig.\ref{fig3} for the samples
with $\eta^{-1} >1$. This figure indicates that, due to the EWPD constraint,
$\sin\theta_L\lesssim0.3$ which is independent of the Higgs data.
Also note that the EWPD constraint is tighter than the $V_{tb}$ and $R_b$ constraints, which require
$|\sin\theta_L|<0.59$ and $|\sin\theta_L|<0.52$ respectively \cite{min-dilaton1}.

%%Fig.4 %%%%%%%%%%%%%%%%%%%%
\begin{figure}[]
\includegraphics[width=15.0cm]{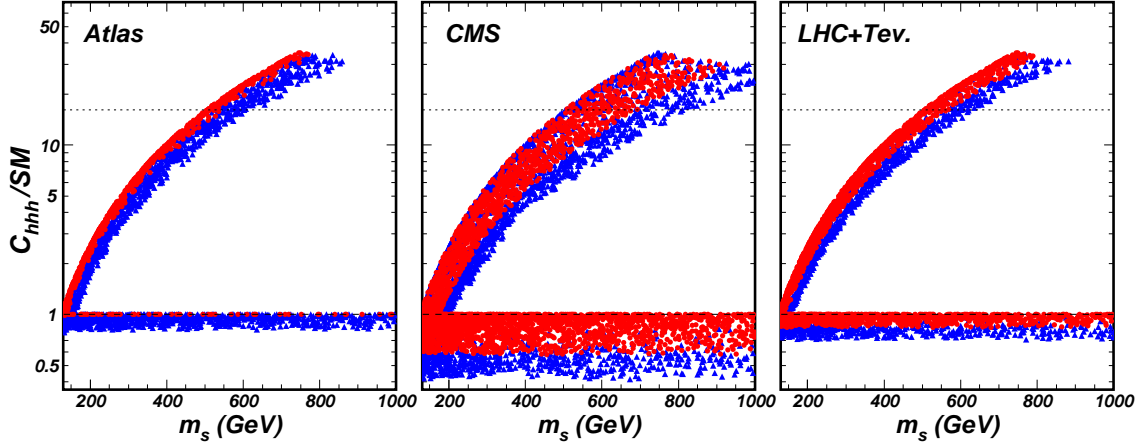}\vspace{0.1cm}
\vspace*{-0.5cm}
\caption{Same as Fig.\ref{fig2}, but projected on the plane of $C_{hhh}/SM$ versus $M_s$.
The dotted line denotes the constraint from perturbativity \cite{perturb} which corresponds to $C_{hhh}/v <4\pi$ (or $C_{hhh}/SM \lesssim 16$).
}
\label{fig4}
\end{figure}
%%%%%%%%%%%%%%%%%%%%
In Fig.\ref{fig4} we project the samples with $\eta^{-1} >1 $ on the plane of $C_{hhh}/SM$ versus $m_s$.
One can learn that, for most of the samples in all the three fits, the Higgs triple self coupling can be either
around the SM value or monotonically increase as $m_s$ goes up. We checked that for samples with greatly enhanced Higgs
self coupling, they are characterized by $0<\tan\theta_S<1$. In this case, $\lambda_H$ may be very large for
heavy dilaton, reaching about 18 in optimum case. We also checked that $\lambda_S$ may also be quite large (reaching 35),
which occurs for $\tan\theta_S<0$, small $\eta^{-1}\equiv f/v$ and large $m_s$. However, due to the $\sin^3 \theta_S$ suppression in Eq(\ref{selfcoup}),
its influence on $C_{hhh}$ is not significant.

Fig.\ref{fig4} indicates that, given $\tan \theta_S > 0$ for the samples passing the ATLAS data, $m_s \gtrsim 800 \GeV$ has
been excluded by the Landau pole constraint. We checked that this upper bound can be further suppressed if we impose
the constraint at a higher scale. Also note that the perturbativity requirement on the Higgs triple coupling,
which corresponds to $C_{hhh}/v <4\pi$ (or $C_{hhh}/SM \lesssim 16$) \cite{perturb}, can set a stronger bound on
$m_s$.

\subsection{Higgs pair production and its detection at the hadron collider}
%%Fig.5 %%%%%%%%%%%%%%%%%%%%
\begin{figure}[]
\includegraphics[width=15.0cm]{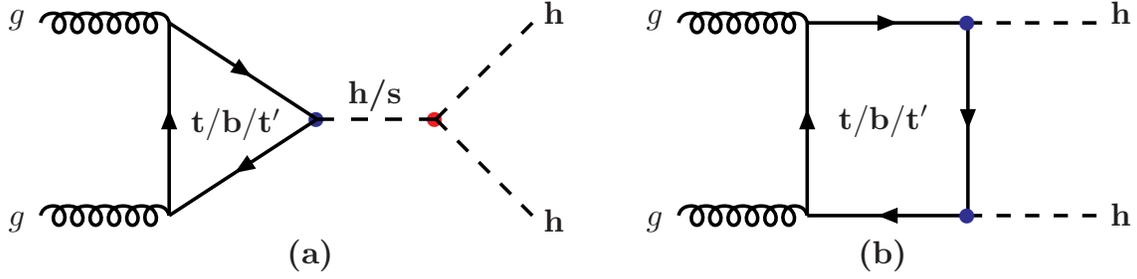}\vspace{0.1cm}
\vspace*{-0.5cm}
\caption{Feynman diagrams for Higgs pair production through gluon fusion at the proton-proton hadron colliders.
}
\label{fig5}
\end{figure}
%%%%%%%%%%%%%%%%%%%%
From the above analysis we have seen that,
the latest $125~{\rm GeV}$ Higgs data along with EWPD have powerful constraints on the Higgs couplings to SM particles. Especially in the fit to the combined data of LHC and Tevatron, the explanation given by MDM is not much better than the SM.
The Higgs triple self coupling, however, can be much larger than the SM value, and we therefore investigate its effect on the Higgs pair production in the following.

We take the $1\sigma$ samples for the combined data as an example, and calculate the Higgs pair production cross section at proton-proton colliders with the modified code $\textsf{HPAIR}$ \cite{hpair}. The relevant Feynman diagrams are shown in Fig.\ref{fig5}, where (a) and (b) correspond to triangular and box diagrams, respectively.
There also exists tree level diagram $b\bar{b}\to hh$, but its contribution is negligible due to the low component of $b$ quark in proton-proton colliders, and to be consistent with the processes generated in the following Monte Carlo (MC) simulation we do not consider it here.
In the SM and many new physics models, the contributions from box diagrams are dominated
, and the top quark contribution is much larger than that from $b$ quark \cite{hh-selfcoup, hh-old, hh-SUSY, hh-2hdm, hh-LH, hh-EDM}.
In the MDM we checked that, the top quark contribution is still much larger than that from the $b$ quark and top partner.
The triangular diagrams, however, may contribute more than the box diagrams due to the possible large Higgs triple self coupling.

%%Fig.6 %%%%%%%%%%%%%%%%%%%%
\begin{figure}[]
\includegraphics[width=15.0cm]{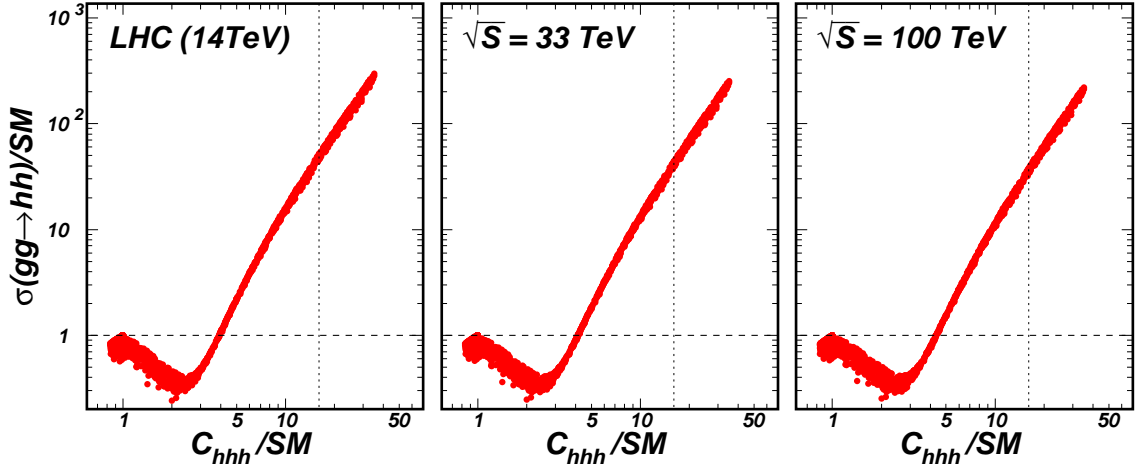}\vspace{0.1cm}
\vspace*{-0.5cm}
\caption{Scatter plot of the samples with $\eta^{-1} > 1$ and passing the LHC+Tevtron data at $1\sigma$ level, projected
on the plane of $\sigma(gg\to hh)/SM$ versus $C_{hhh}/SM$. $\sigma(gg\to hh)/SM$ is the 125-GeV Higgs pair production rates
without cuts, calculated at LHC14, TeV33 and TeV100 proton-proton colliders, respectively.}
\label{fig6}
\end{figure}
%%%%%%%%%%%%%%%%%%%%
We calculate the Higgs pair production cross sections at proton-proton colliders with $\sqrt{S}=14, ~33, ~100 \TeV$ in both  MDM and SM. In Fig.\ref{fig6}, the $1\sigma$ samples for the combined data are projected on the plane of the cross section rate $\sigma(gg\to hh)/SM$ versus the self coupling rate $C_{hhh}/SM$.
We can see that with the self coupling increasing, the cross section rates first monotonically reduce then monotonically increase at all these collider energies, and the turning point is at $C_{hhh}/SM\simeq2.5$.
The reason is that the total amplitude of the triangle diagrams has an opposite sign compared to that from the box diagrams,
and in SM the contribution from the box diagrams is dominant.
Thus for $C_{hhh}/SM\gtrsim2.5$ the contribution from triangle diagrams becomes dominant and monotonically increases when the self coupling goes up.
At 14-TeV LHC the production rate can be as large as about $50$ for $C_{hhh}/SM\simeq16$ or $m_s\simeq500\GeV$,
and at 33 TeV and 100 TeV colliders it can also reach $30$.

Since the Higgs pair production rate in the heavy dilaton scenario can be highly enhanced,
we can expect its discovery at the proton-proton colliders with a moderate or small integral luminosity.
The author of \cite{hhmc-Yao} performed a detailed MC simulation of the Higgs pair production in the SM,
through $gg\to hh\to b\bar{b}\gamma\gamma$ at proton-proton colliders with a high integral luminosity of $3000\fbm$.
In this work, we use the result of \cite{hhmc-Yao} in the SM but reduce the integral luminosity from $3000\fbm$ to $100\fbm$.
Table \ref{MCtable} is taken from \cite{hhmc-Yao} for the convenience of discussion, but the number of expected events are modified according to the reduction of integral luminosity from $3000\fbm$ in \cite{hhmc-Yao} to $100\fbm$ here.
We assume that in the MDM the $\sigma\times Br$ and acceptances of the background,
the acceptances of the signal are the same as that in the SM,
while the $\sigma\times Br$ of the signal are calculated by ourselves.

\begin{table}
\caption{$\sigma\times Br$, acceptance, and the expected events of the signal and background processes at an integral luminosity of $100\fbm$ for the proton-proton colliders with $\sqrt{s}=14, ~33 ~{\rm and}~ 100 \TeV$. The table is taken from \cite{hhmc-Yao} for the convenience of discussion, but the number of expect events are modified according to the reduction of integral luminosity from $3000\fbm$ in \cite{hhmc-Yao} to $100\fbm$ here.}
\label{MCtable}
\vspace{.3cm}
\begin{tabular}{|c|ccc|ccc|ccc|}
\hline \hline
Samples
& \multicolumn{3}{c}{LHC14 ($100\fbm$) }
& \multicolumn{3}{|c}{TeV33 ($100\fbm$)}
& \multicolumn{3}{|c|}{TeV100 ($100\fbm$)}
\\
\hline
   & $\sigma \cdot Br$ & Acc. & Expect
   & $\sigma \cdot Br$ &  Acc. &  Expect
   &$\sigma \cdot Br$ &  Acc. & Expect
\\
\hline
   & (fb) & (\%) & Events
   & (fb) & (\%) & Events
   & (fb) & (\%) & Events
\\
\hline
hh($b\bar b \gamma\gamma$)
& 0.089 & 6.2  & 0.552
& 0.545 & 5.04 & 2.75
& 3.73  & 3.61 & 14.47
\\
\hline
$b\bar b \gamma\gamma$
&294  &0.0045  &1.323
&1085 &0.0039  &4.23
&5037 &0.00275 &13.85
\\
$z(b\bar b) h(\gamma\gamma)$
&0.109 &1.48 &0.161
&0.278 &1.41 &0.392
&0.875 &1.57 &1.374
\\
$b\bar b h(\gamma\gamma)$
&2.23 &0.072 &0.161
&9.84 &0.084 &0.827
&50.5 &0.099 &5.00
\\
$t\bar t h(\gamma\gamma)$
&0.676 &0.178 &0.1203
&4.76  &0.12  &0.571
&37.3  &0.11  &4.103
\\
\hline
Total B
&-- &-- & 1.765
&-- &-- & 6.02
&-- &-- & 24.33
\\
\hline
S/$\sqrt{B}$
&-- &-- & 0.42
&-- &-- & 1.12
&-- &-- & 2.93
\\
\hline

\hline \hline
\end{tabular}
\end{table}

%%Fig.7 %%%%%%%%%%%%%%%%%%%%
\begin{figure}[]
\includegraphics[width=15.0cm]{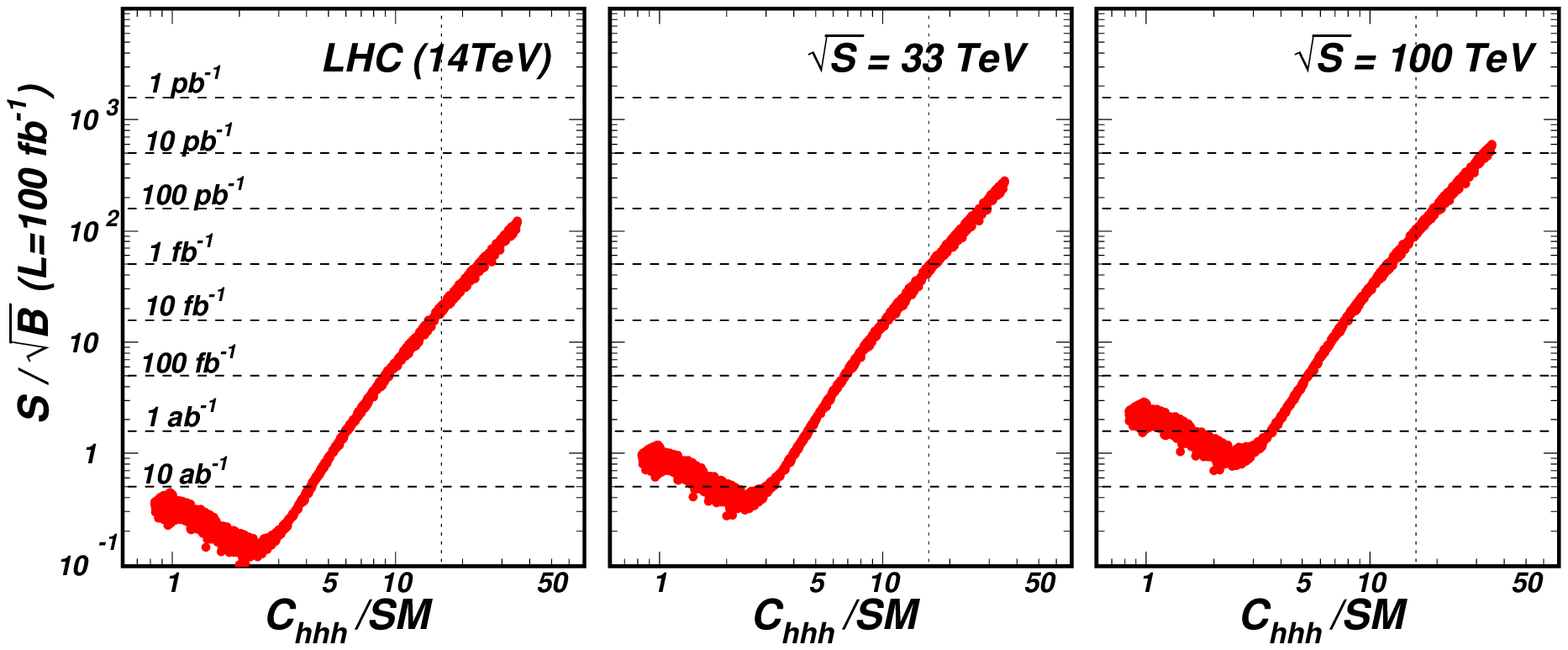}\vspace{0.1cm}
\vspace*{-0.5cm}
\caption{Same as Fig.\ref{fig6}, but projected on the planes of $S/\sqrt{B}$ versus $C_{hhh}/SM$.
The $S/\sqrt{B}$ is calculated at the integral luminosity of $100\fbm$,
while $S/\sqrt{B}=5$ for other value of luminosity are also marked out,
which are the discovery limits for the corresponding luminosity.
}
\label{fig7}
\end{figure}
%%%%%%%%%%%%%%%%%%%%

In Fig.\ref{fig7}, we project the $1\sigma$ samples for the combined data on the plane of significance $S/\sqrt{B}$ versus coupling $C_{hhh}/SM$, where $S/\sqrt{B}$ is calculated with an integral luminosity of $100\fbm$. $S/\sqrt{B}=5$ for other values of luminosity are also marked out, which are the discovery limits for the corresponding luminosity.
We can see that, higher energy will make the proton-proton colliders more capable of detecting the signal of Higgs pair production.
In addition, when $C_{hhh}/SM\gtrsim2.5$ ($m_s\gtrsim 200$ GeV) and goes up, the significance $S/\sqrt{B}$ increases monotonically.
For example, for $C_{hhh}/SM=16$ or $m_s=500\GeV$, the discovery luminosity is about $10, ~1, ~0.5 \fbm$ for LHC14, TeV33 and TeV100, respectively.
And with luminosity of $100\fbm$ at LHC14, the Minimal Dilaton Model with $C_{hhh}/SM\simeq9$ or $m_s=400\GeV$ may be covered.

\section{Higgs phenomenology in the light dilaton scenario}

For a light dilaton with mass $m_s<m_h/2\simeq 62\GeV$, the most interesting phenomenology is the Higgs exotic decay $h\to ss$. Similar exotic decay is also very attractive in Supersymmetric models and Little Higgs Models
\cite{Light-SUSY, Light-SLH}, such as in the Next-to-Minimal Supersymmetric Standard Model (NMSSM) according to the latest paper \cite{Light-1309}.
In the following, we define samples with $\Delta\chi^2<1$, $1<\Delta\chi^2<4$ and $4<\Delta\chi^2<9$ to be $1\sigma$, $2\sigma$ and $3\sigma$ samples respectively,
where $\Delta\chi^2=\chi^2-\chi^2_{min}$ and $\chi^2_{min}$ denote the same global minimal values in the three fits as those in the heavy dilaton scenario.
For these samples, if one projects them on the plane of any observable $O_i$ versus $\Delta \chi^2$, one can get the allowed ranges of $O_i$ at $1\sigma$, $2\sigma$ and $3\sigma$ level, respectively.  Note that the classification standard here is different from that in the heavy dilaton scenario discussed in the previous sections.

The strategies we take in the following are similar to those in the heavy dilaton scenario, except for:
\begin{itemize}
\item [(i)] The range of $m_s$ is shifted from $130<m_s<1000\GeV$ to $0<m_s<62\GeV$;
\item [(ii)]
Classification standard of $n\sigma$ for the samples is changed according to the above description;
  \item [(iii)]
As an example, analysis of the samples for the CMS data is paid more attention and $\eta$-fixed fits are performed and investigated.
\end{itemize}
Besides, we remark that since the couplings $\lambda_H$, $\lambda_S$ and $\kappa$ are small or moderate in light dilaton scenario,
the Landau pole constraints and perturbation requirement can not impose any limitation.

\subsection{The light dilaton scenario confronts with the current Higgs data}

%%Fig.8 %%%%%%%%%%%%%%%%%%%%
\begin{figure}[]
\includegraphics[width=15.0cm]{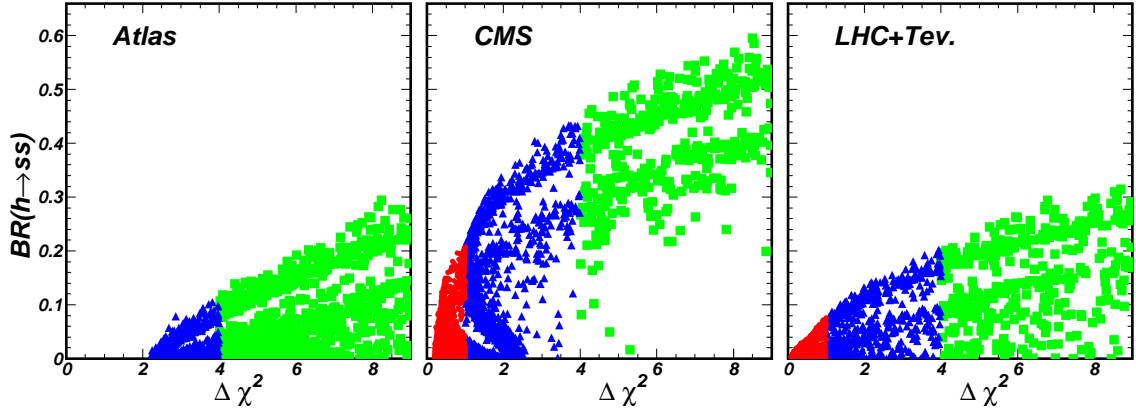}\vspace{0.1cm}
\vspace*{-0.5cm}
\caption{The scatter plots of the samples satisfying constraints from the ATLAS data (left),
the CMS data (middle) and the LHC+Tevtron data (right), respectively, projected on the plane
of $BR(h \to ss)$ versus $\Delta\chi^2=\chi^2-\chi^2_{min}$. All the samples have passed constraints
from the EWPD and the Higgs search from LEP, Tevatron and LHC. Samples with $\Delta\chi^2 <1$ (red bullets),
$1< \Delta\chi^2 <4$ (blue triangles) and $4< \Delta\chi^2 <9$ (green square) are called $1\sigma$,
$2\sigma$ and $3\sigma$ samples, respectively.}
\label{fig8}
\end{figure}
%%%%%%%%%%%%%%%%%%%%
To classify the samples in the light dilaton scenario, we first project the samples on the plane of $BR(h \to ss)$ versus $\Delta\chi^2$ in Fig.\ref{fig8}.
This figure shows that, the exotic decay $h \to ss$ for the CMS data can have the largest branching ratio,
which can reach $21\%$, $43\%$ and $60\%$ at $1\sigma$, $2\sigma$ and $3\sigma$ level, respectively.
This is because most of the inclusive signal rates of the CMS data are suppressed compared to the SM values, thus
a relatively large exotic branching ratio is more favored.
On the contrary, samples for the ATLAS data have smaller exotic branching ratios and very interestingly, there is no $1\sigma$ samples at all.
The branching ratio for combined data, however, is at most about $32\%$ at $3\sigma$ level, which is slightly larger than the value allowed in SM.

%%Fig.9 %%%%%%%%%%%%%%%%%%%%
\begin{figure}[]
\includegraphics[width=15.0cm]{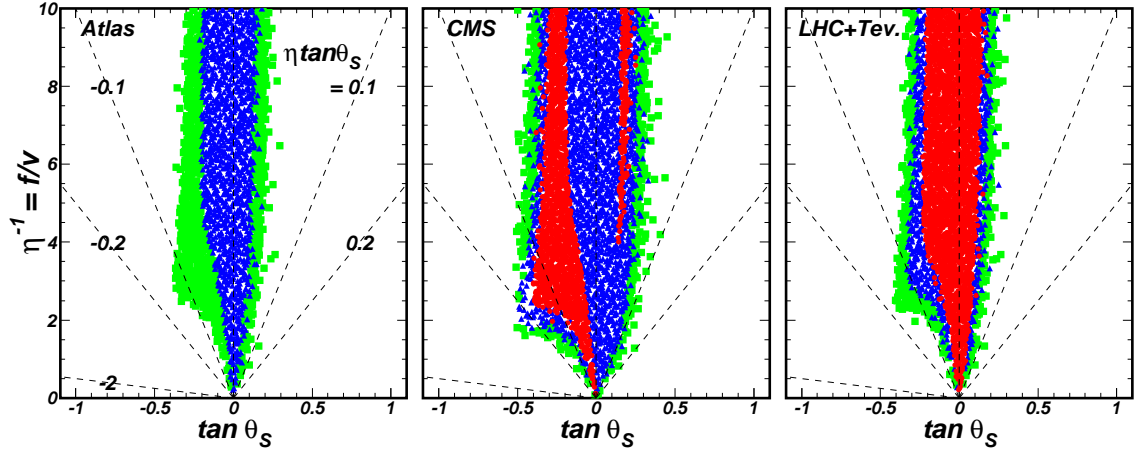}\vspace{0.1cm}
\vspace*{-0.5cm}
\caption{Same as Fig.\ref{fig8}, but projected on the plane of $\eta^{-1}\equiv f/v$ versus $\tan\theta_S$. Dashed lines of $\tan\theta_S=0$ and $\eta\tan\theta_S=\pm0.1,\ \pm0.2, -2$ are also shown,
which correspond to the SM point and lines of $(C_{hgg}/SM)/(C_{hff}/SM)=1.1,\ 0.9,\ 1.2,\ 0.8, -1$
or $(C_{h\gamma\gamma}/SM)/(C_{hff}/SM)=0.973,\ 1.027,\ 0.945,\ 1.055,\ 1.55$ in Fig.\ref{fig2}, respectively.
}
\label{fig9}
\end{figure}
%%%%%%%%%%%%%%%%%%%%
To intensively study the light dilaton scenario and also compare with the heavy dilaton scenario,
in Fig.\ref{fig9} we project the samples on the plane of $\eta^{-1}\equiv f/v$ versus $\tan\theta_S$ again.
We would like to emphasize again that the classification standard here is different from that in the heavy dilaton scenario.
Combining Fig.\ref{fig9} with Fig.\ref{fig1}, Fig.\ref{fig2} and Fig.\ref{fig8}, we can see that:
\begin{itemize}
\item In the light dilaton scenario, the surviving samples are mostly confined in $|\eta\tan\theta_S|<0.2$
and there are no samples with $\eta\tan\theta_S\simeq-2$. Compared with the corresponding results of
the heavy dilaton scenario, the Higgs data have stronger constrains on the
parameter space in the light dilaton scenario.
\item Due to the narrower range of $|\eta\tan\theta_S|$, the various Higgs couplings in the light dilaton
scenario can not deviate much from the SM value. For example, most of the $2\sigma$ samples for the ATLAS
and combined data have $|(C_{hgg}/SM)/(C_{hff}/SM)-1|<0.1$.
\item The CMS data have interesting constrains on the light dilaton scenario. The region of $1\sigma$
samples has two separate parts with the broader one characterized by a negative $\tan\theta_S$ and the
narrow one corresponding $\eta^{-1}\gtrsim4$.
\end{itemize}

Since the samples for the CMS data can accommodate larger branching ratio of exotic Higgs decay
in the light dilaton scenario, we will concentrate on it from now on. We perform three further scans with fixed $\eta^{-1}=1,~2.5,~5$, respectively. Then for the samples satisfying the EWPD and light Higgs search data,
we calculate the $\chi^2$ using the CMS data only, and classify the samples into $1\sigma$, $2\sigma$ and $3\sigma$ region as done in this sector.

%%Fig.10 %%%%%%%%%%%%%%%%%%%%
\begin{figure}[]
\includegraphics[width=15.0cm]{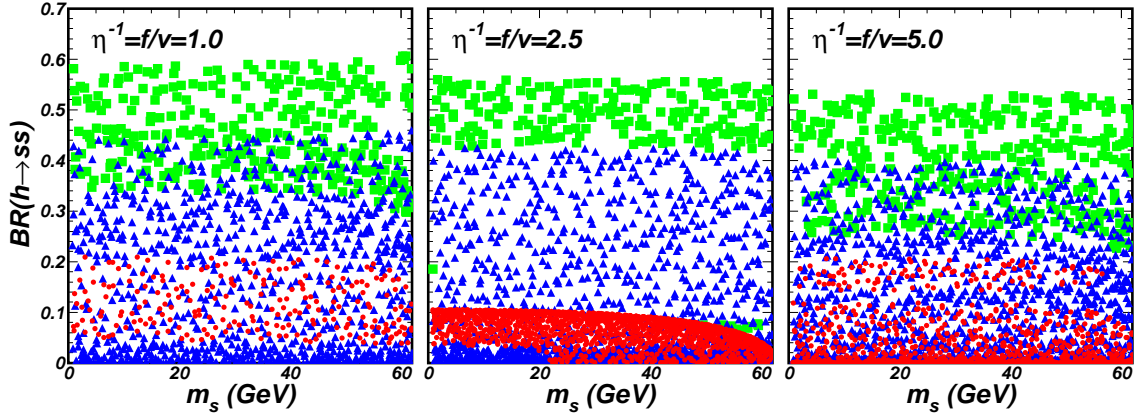}\vspace{0.1cm}
\vspace*{-0.5cm}
\caption{The scatter plots of the samples satisfying EWPD and the CMS data, projected on the plane of $BR(h \to ss)$ versus $m_s$.
The samples are classified as in Fig.\ref{fig8}, but with $\eta^{-1}\equiv f/v$ fixed at 1 (left), 2.5 (middle) and 5 (right), respectively.
}
\label{fig10}
\end{figure}
%%%%%%%%%%%%%%%%%%%%
In Fig.\ref{fig10}, we project the $\eta$-fixed samples for the CMS data on the plane of $BR(h \to ss)$ versus $m_s$.
This figure shows that the upper bound of the rare decay is roughly independent of the light dilaton mass $m_s$. The
reason is that $m_s$ as an input parameter of the model is scarcely limited by the constraints we considered, and meanwhile it only
affect the rate by phase space. As a consequence, the rare decay rate is mainly determined by the $C_{hss}$ or basically,
by $\eta^{-1}$ and $\tan\theta_S$ (see Eqs.(\ref{lambda}) and (\ref{hsscoup})).

%%Fig.11 %%%%%%%%%%%%%%%%%%%%
\begin{figure}[]
\includegraphics[width=15.0cm]{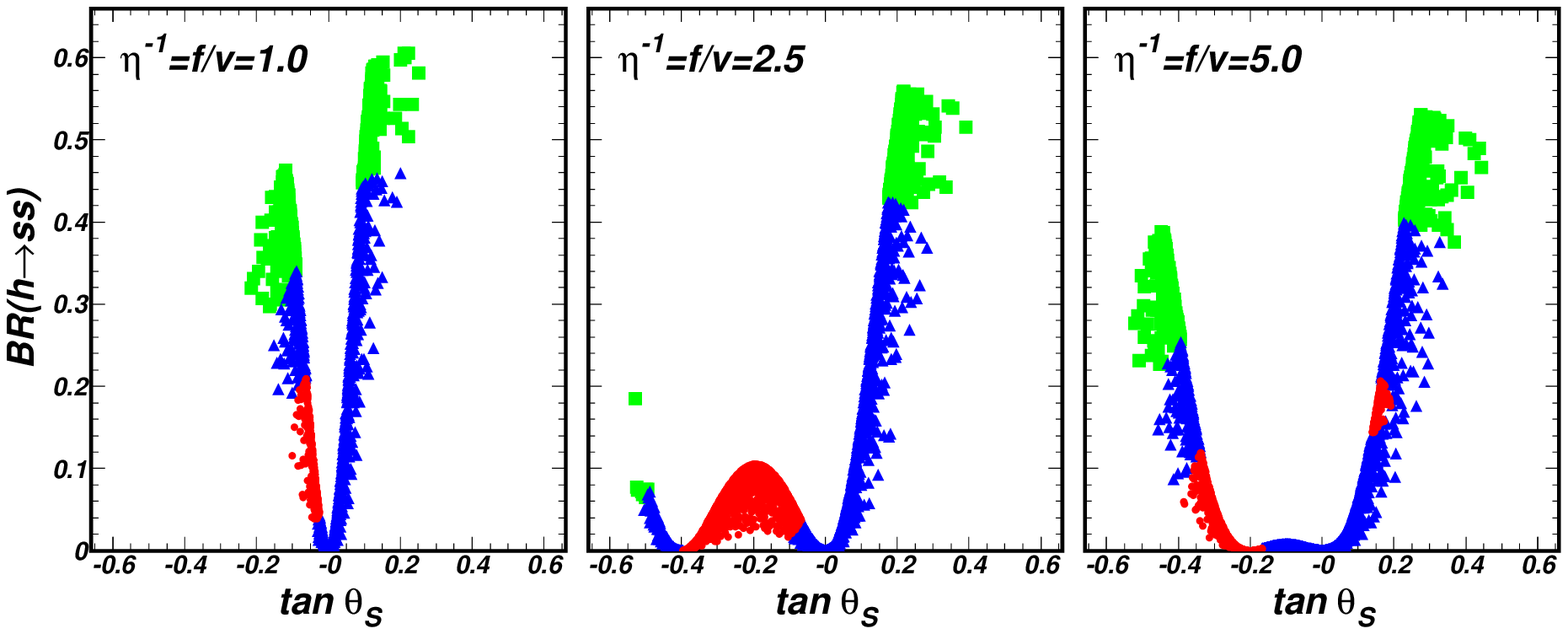}\vspace{0.1cm}
\vspace*{-0.5cm}
\caption{Same as Fig.\ref{fig10}, but projected on the plane of $BR(h \to ss)$ versus $\tan\theta_S$.
}
\label{fig11}
\end{figure}
%%%%%%%%%%%%%%%%%%%%
In Fig.\ref{fig11}, we project the samples on the plane of $BR(h \to ss)$ versus $\tan\theta_S$.
Combining with Fig.\ref{fig9}, we can see that:
\begin{itemize}
  \item
For a small $\eta^{-1}\equiv f/v$, e.g., $\eta^{-1}=1$,
the $h \to ss$ branching ratio can reach $60\%$ while $\theta_S$ is confined in a narrow region of $|\tan\theta_S|\lesssim0.2$.
This is because a large $|\tan\theta_S|$ with a large $\eta$ will modify the Higgs couplings and signal rates too much,
which is disfavored by the current Higgs data.
  \item
For a moderate $\eta^{-1}\equiv f/v$, e.g., $\eta^{-1}=2.5$,
all of the $1\sigma$ samples have negative $\tan\theta_S$ and small $Br(h \to ss)$,
while most of the samples with large $Br(h \to ss)$ have positive $\tan\theta_S$.
Both of these two features can be understood from $(C_{hgg}/SM)/(C_{hff}/SM)=1+\eta\tan\theta_S$. Considering that the CMS data favor suppressed signal rates, which means $(C_{hgg}/SM)/(C_{hff}/SM)$ should be around 1 or even less. If $\eta\tan\theta_S$ is too large, then a large exotic decay width would be needed to enhance the total width in order to suppress various signal rates. If $\eta\tan\theta_S$ is negative, however, a small $Br(h \to ss)$ would be enough.
  \item
For a large $\eta^{-1}\equiv f/v$, e.g., $\eta^{-1}=5$,
the $1\sigma$ samples begin to appear in the $\tan\theta_S>0$ region,
which is the result of the interplay between the exotic branching ratio and Higgs couplings.
We can make an estimation of the inclusive $pp\to h\to XX$ signal rate in the $\tan\theta_S>0$ region, which is approximately
\begin{eqnarray}
  R_{XX}
  &\simeq& [(C_{hgg}/SM)/(C_{hff}/SM)]^2 \times [1-BR(h \to ss)] \times (C_{hXX}/SM)^2 \nonumber \\
  &\simeq& (0.85\sim0.90) \times (C_{hXX}/SM)^2 \nonumber \\
  &\simeq&\Big\{
\begin{array}{cc}
  (0.83\sim0.88), & ~~{\rm for}~ XX=VV^*, ff \\
  (0.76\sim0.81), &   {\rm for}~ XX=\gamma\gamma~~~~~~
\end{array}
\end{eqnarray}
We can see that $\tan\theta_S>0$ region can accommodate $1\sigma$ samples for the CMS data, of which the inclusive signal rates of $\gamma\gamma$, $ZZ^*\to 4\ell$ are $0.77\pm 0.27$, $0.92\pm0.28$, respectively \cite{1303-c-com}.
\end{itemize}

\subsection{Detection of light dilaton at hadron and lepton colliders}

From the above analysis, we see that in the light dilaton scenario the Higgs couplings to SM particles can not deviate much from their SM values, and consequently the dilaton couplings to SM particles should be very small according to Eqs.(\ref{scoupling}) and (\ref{hcoupling}).
Thus it will be very hard to detect the light dilaton from its couplings to SM particles,
such as through its associated production with $Z$ boson at hadron or lepton colliders.
Nevertheless, since the Higgs exotic decay $h \to ss$ can have a large branching ratio according to the CMS data at $3\sigma$ level, it may be possible to detect the light dilaton at LHC14 or Higgs factory in the future.

%%Fig.12 %%%%%%%%%%%%%%%%%%%%
\begin{figure}[]
\includegraphics[width=15.0cm]{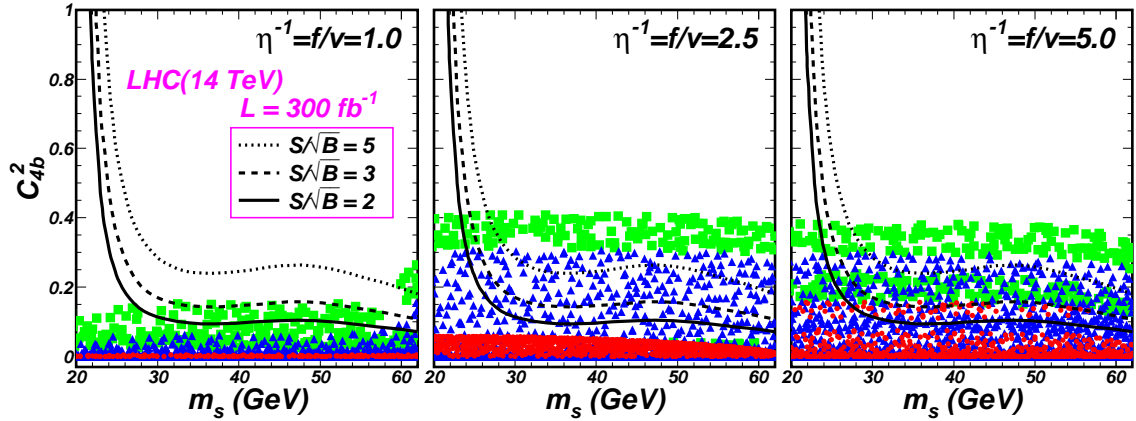}\vspace{0.1cm}
\vspace*{-0.5cm}
\caption{Same as Fig.\ref{fig10}, but projected on the plane of $C_{4b}^2$ versus $m_s$. The significance curves of $S/\sqrt{B}=2,~3,~5$ are taken from \cite{Light-1309} for an intergraded luminosity of $300 \fbm$ at the LHC ($14 ~{\rm TeV}$).
}
\label{fig12}
\end{figure}
%%%%%%%%%%%%%%%%%%%%
To detect the light dilaton through $pp\to hZ, ~h \to ss \to 4b$ at LHC14,
we use the result of MC simulation performed at LHC14 in our previous work \cite{Light-1309} with a luminosity of $300\fbm$.
The setting of $\sigma\times Br$ and acceptances are the same as in the heavy dilaton scenario.
The $\eta$-fixed samples for the CMS data are projected on the plane of $C_{4b}^2$ versus $m_s$ in Fig.\ref{fig12}, where the significance curves of $S/\sqrt{B}=2,~3,~5$ are taken from \cite{Light-1309}. The signal rate $C^2_{4b}$ is defined as
\begin{eqnarray}
  C^2_{4b} &=& (C_{hVV}/SM)^2 \times BR(h \to ss) \times [BR(s\to b\bar{b})]^2.
\end{eqnarray}
From this figure we can see that:
\begin{itemize}
  \item
For $\eta^{-1}=2.5$ and 5.0, basically only the $3\sigma$ samples can be possibly detected, and most of the other samples are below the curve of $S/\sqrt{B}=5$ which means larger luminosity is needed to detect the $1\sigma$ and $2\sigma$ samples.
  \item
For $\eta^{-1}=1$, most samples are with $C^2_{4b}\lesssim0.2$,
which are out of the detection capability of LHC14 with a luminosity of $300\fbm$.
This is because most of the $\eta^{-1}=1$ samples have $|\tan\theta_S|\lesssim0.2$,
which means the dilaton coupling to $b$ quark is limited to be $|C_{sbb}/SM|= |-\sin\theta_S| \lesssim 0.2$.
Therefore the branching ratios of the light dilaton satisfy $BR(s\to b\bar{b})\ll BR(s\to gg)\simeq1$ and result in a small $C^2_{4b}$.
We also checked that for samples with $\eta^{-1}\lesssim1$, $C^2_{4b}$ are all very small, which means light dilaton with $\eta^{-1}\equiv f/v \lesssim1$ can not be detected at the LHC14 through $pp\to hZ \to 4bZ$ with a luminosity of $300\fbm$.
  \item
Combining with Fig.\ref{fig11} we also checked that, samples with
(i) $\theta_S$ around $0$, e.g. $|\tan\theta_S|\lesssim0.2$,
(ii) small $m_{s}\lesssim 2m_{b}\simeq10 \GeV$,
and (iii) small exotic branching ratio, e.g. $BR(h \to ss)\lesssim0.2$,
also have rather small $C^2_{4b}$ which can not be detected with a luminosity of $300\fbm$.
\end{itemize}

%%Fig.13 %%%%%%%%%%%%%%%%%%%%
\begin{figure}[]
\includegraphics[width=15.0cm]{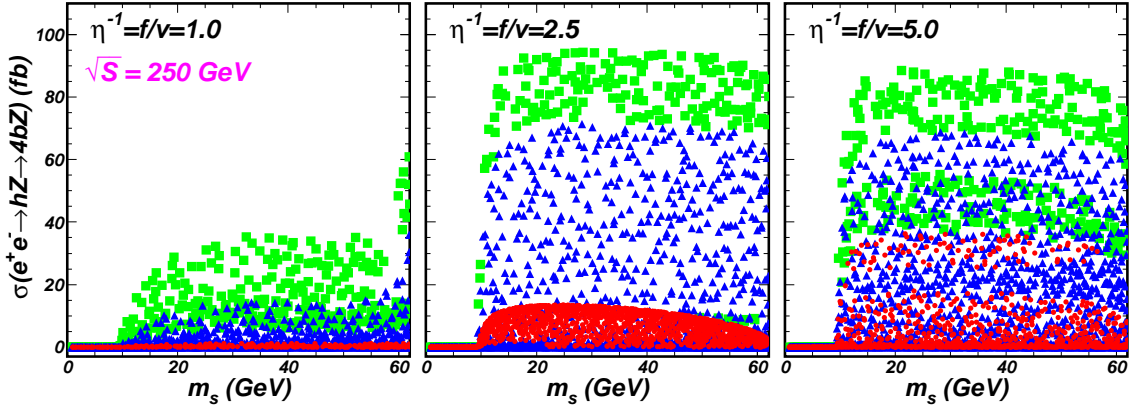}\vspace{0.1cm}
\vspace*{-0.5cm}
\caption{Same as Fig.\ref{fig10}, but projected on the planes of $\sigma(e^{+}e^{-}\to hZ \to 4bZ)$ versus $m_s$.
}
\label{fig13}
\end{figure}
%%%%%%%%%%%%%%%%%%%%
Similar to \cite{Light-1309}, we also investigated the detection of light dilaton at the electron-positron collider through $e^+e^-\to hZ, ~h \to ss \to 4b$.
The cross sections are calculated for a collision energy of $\sqrt{S}=250\GeV$, and
the result is shown on the plane of $\sigma(e^+e^-\to hZ \to 4bZ)$ versus $m_s$ in Fig.\ref{fig13}.
We can see that, for samples with $\eta^{-1}\gtrsim2$ and $m_s\gtrsim10\GeV$,
the cross section can reach $80\fb$ at $3\sigma$ level, and for $1\sigma$ samples with $\eta^{-1}=5$ it can still reach $30\fb$.
Thus we can expect that it would be easier to detect the light dilaton at the electron-positron collider at $\sqrt{S}=250\GeV$ with its much cleaner background compared to the hadron colliders.
Larger collision energy, however, will result in smaller cross section due to the s channel property of the $hZ$ associated production. And a large $Br(h \to ss)$ and not-too-small $C_{sbb}$, or a large $\eta^{-1}\equiv f/v$ and not-too-small $|\tan\theta_S|$, is favorable for the detection of light dilaton at electron-positron colliders as well as hadron colliders.

\section{conclusion}

In this work, we consider the theoretical and experimental constrains on the MDM, such as the vacuum stability,
the absence of the Landau pole,  the EWPD and the Higgs search at the LEP, Tevatron and LHC.  Then we perform
 fits to the latest 125-GeV Higgs data both in the heavy dilaton scenario and in the light dilaton scenario.
Noting the apparent difference between the ATLAS and CMS data,
in our fits we consider the ATLAS data, the CMS data and the combined data of LHC and Tevatron separately. For each scenario,
we consider following aspects:
\begin{itemize}
  \item [(1)]
In the heavy dilaton scenario, we show the surviving parameter space and various Higgs couplings to SM particles
for the $1\sigma$ and $2\sigma$ samples in the fits. Modification of the Higgs triple self coupling is also discussed.
For the $1\sigma$ samples fitted to the combined data,
we calculate the Higgs pair production cross section at the proton-proton colliders such as LHC14 ($\sqrt{S}=14\TeV$), TeV33 ($\sqrt{S}=33\TeV$) and TeV100 ($\sqrt{S}=100\TeV$), and discuss the deviations from the SM values.
Based on the MC simulation result in \cite{hhmc-Yao}, we also investigate the significance $S/\sqrt{B}$ at these colliders.
  \item [(2)]
In the light dilaton scenario, we first show the ranges of exotic decay branching ratio $Br(h \to ss)$ in different fits,
and compare the surviving parameter space with that in the heavy dilaton scenario.
Then we fix $\eta^{-1}$ to perform fits only to the CMS data with particular attention paid to the
dependence of $BR(h \to ss)$ on $\eta^{-1}$ and $\tan\theta_S$.
Based on these $\eta^{-1}$-fixed samples, we study the detection of light dilaton through $pp\to hZ \to 4bZ$ at the LHC14 with a luminosity
of $300\fbm$ by our previous MC simulation \cite{Light-1309}. We also discuss the detection at the electron-positron collider with $\sqrt{S}=250\GeV$.
\end{itemize}

Finally we have following observations
\begin{itemize}
\item
If one considers the ATLAS and CMS data separately, the MDM can explain each of them well,
but refer to different parameter space due to the apparent difference in the two sets of data.
If one considers the combined data of the LHC and Tevatron, however, the explanation given by the MDM is not much better than the SM,
and the dilaton component in the 125-GeV Higgs is less than about $20\%$ at $2\sigma$ level.
\item
The current Higgs data have stronger constrains on the light dilaton scenario than the heavy dilaton scenario.
\item
The heavy dilaton scenario can produce a Higgs triple self coupling much larger than the SM value, and thus a significantly enhanced Higgs pair cross section at hadron colliders.
With a luminosity of $100\fbm$ ($10\fbm$) at the 14-TeV LHC, a heavy dilaton of $400\GeV$ ($500\GeV$) can be examined.
\item
In the light dilaton scenario, the Higgs exotic branching ratio can reach $43\%$ (60\%) at $2\sigma$ ($3\sigma$) level when considering only the CMS data, which may be detected at the 14-TeV LHC with a luminosity of $300\fbm$ and the Higgs Factory.
\end{itemize}

\section*{Acknowledgement}
We thank Yan-Yan Bu, Cheng Li for helpful discussion on the Minimal Dilaton Model.
This work was supported in part by the National Natural
Science Foundation of China (NNSFC) under grant No. 10775039, 11075045,
11275245, 10821504 and 11135003, by Program for New Century Excellent Talents in University,
and by the Project of Knowledge Innovation Program (PKIP) of Chinese Academy of Sciences under grant No. KJCX2.YW.W10.

{\em Note added:}
When this manuscript is being prepared, a paper on a similar subject but with much different content appeared in the arXiv \cite{h-dilaton-mix}.

\end{document}